\documentclass{iau}
\usepackage{epsf}
 
\def\etal{{et al.~}}
\def\Msun{\> M_{\odot}}
\def\kpc{\> {\rm kpc}}
\def\Gyr{\> {\rm Gyr}}
\newcommand{\kms}{\>{\rm km}\,{\rm s}^{-1}}        

\title{Local Group Proper Motion Dynamics}

\author[van der Marel]{Roeland P.~van der Marel$^1$} 

\affiliation{$^1$Space Telescope Science Institute, 3700 San Martin Drive,
Baltimore, MD 21218}

\pubyear{2014}
\volume{311}
\jname{Galaxy Masses as Constraints of Formation Models}
\editors{M. Cappellari \& S. Courteau, eds.}

\begin{document}

\maketitle

\begin{abstract}
Our knowledge of the dynamics and masses of galaxies in the Local
Group has long been limited by the fact that only line-of-sight
velocities were observationally accessible. This introduces
significant degeneracies in dynamical models, which can only be
resolved by measuring also the velocity components perpendicular to
the line of sight. However, beyond the solar neighborhood, the
corresponding proper motions have generally been too small to
measure. This has changed dramatically over the past decade,
especially due to the angular resolution and stability available on
the Hubble Space Telescope. Proper motions can now be reliably
measured throughout the Local Group, as illustrated by, e.g., the work
of the HSTPROMO collaboration. In this review, I summarize the
importance of proper motions for Local Group science, and I describe
the current and future observational approaches and facilities
available to measure proper motions. I highlight recent results on
various Milky Way populations (globular clusters, the bulge, the
metal-poor halo, hypervelocity stars, and tidal streams), dwarf
satellite galaxies, the Magellanic Clouds and the Andromeda System.
\end{abstract}

\firstsection

\section{Introduction: Local Group Dynamics and the Importance of 
Proper Motions}

\smallskip
\noindent Structures in the Universe cluster on various scales. Our
Milky Way Galaxy (MW) belongs to a small group, called the Local Group
(LG). The two dominant galaxies in the LG are spiral
galaxies: the MW and the Andromeda Galaxy (M31). Each of these
galaxies has a significant companion that is roughly one-tenth of its
mass: the MW has the Magellanic Clouds, i.e. the pair of the Large and
Small Magellanic Clouds (LMC and SMC), and M31 has the Triangulum
Galaxy (M33). The MW and M31 also each have their own system of dwarf
satellite galaxies, which have much lower mass.

The Universe evolves hierarchically, with small structures merging and
falling in to form bigger structures. Owing to its proximity, the LG
is the best place to study some of these processes in detail. The last
decade has seen a wealth of new discoveries in this area. For example,
the ongoing disruption of the Sagittarius dwarf spheroidal (Sgr dSph)
galaxy has produced a giant stream of stars around the Milky Way. The
SDSS survey has revealed many other such streams, and there is also a
giant stellar stream around M31. For these reasons, the LG and its two
dominant spirals have become the benchmark for testing many aspects of
galaxy formation and cosmological theories.

To understand the LG, it is necessary to study the dynamics
of its stars and galaxies. The dynamics contain an imprint of the
initial conditions, and also reflect subsequent evolution. Moreover,
dynamical measurements are necessary to constrain the amount of (dark)
mass, since the dynamics, structure, and mass of galaxies are tied
through the laws of gravity and dynamical equilibrium. 

Almost everything that is known about the dynamics of the LG, and of
galaxies in the Universe in general, is based on observations of
line-of-sight (LOS) velocities. Such observations constrain only one
component of motion, and interpretation therefore generally requires
that various assumptions be made. This limits the amount of insight
that can be gained, e.g., through the well-known degeneracy between
mass and velocity dispersion anisotropy (Binney \& Mamon 1987).

An important step forward is to determine fully three-dimensional
velocities, by also measuring proper motions (PMs) in the plane of the
sky. This is possible with several techniques, as discussed in
Section~2. PM measurements over the past decade have started to
revolutionize our understanding of the LG, as reviewed (roughly in
order of increasing distance) in Sections~3--6. Concluding remarks are
presented in Section~7.

\section{Observational Approaches and Future Prospects}

\noindent High PM accuracy is required to meaningfully address most
dynamical topics in the LG. For reference, $\Delta$ PM $= 50 \>
\mu$as/yr corresponds to a velocity accuracy $\Delta v \approx (D/4) $
km/s at distance $D$ kpc. Therefore, this is the accuracy required to
study, e.g., MW halo objects at $40 \kpc$ with $10 \kms$ accuracy, or
to study dwarf satellite galaxies at $200 \kpc$ with $50 \kms$
accuracy.  To properly grasp the challenge that this poses, it is
worth noting that $50 \> \mu$as/yr corresponds roughly to the angular
velocity with which one would observe the hair grow of a human placed
at the distance of the moon.

The highest PM accuracies, $\Delta$ PM $\lesssim 10 \> \mu$as/yr, are
generally obtained using interferometric telescope arrays at radio
wavelengths. However, the number of sources in the LG that can be
studied in this way is small. Water masers are the main targets, and
these exist only in the few LG galaxies with high star formation
rates. PM measurements have been reported for M33 and IC10, both
satellites of M31 (Brunthaler \etal 2005, 2007). In the future, such
measurements may become possible also for, e.g., M31 (Darling 2011)
and the Magellanic Clouds.

Ground-based optical PM measurements have traditionally been based on
comparison of photographic plate data taken at widely separated epochs
(decades apart; e.g., Carlin \etal 2012). A variant of this is to
compare historical photographic plate data with recent targeted CCD
data (e.g., Casetti-Dinescu \etal 2007). Another approach is to
compare homogeneous data from large surveys (e.g., SDSS) using shorter
time baselines (e.g., Koposov \etal 2013). The accuracies thus reached
are comparable to what is obtained from near-IR measurements for stars
near the Galactic Center (e.g., Ghez \etal 2008). However, those stars
move unusually fast around the MW's supermassive black hole, and they
are relatively nearby. The ground-based techniques do not generally
reach PM accuracies {\it on a per-star basis} that are sufficient for
studies of Local Group dynamics. However, by calculating the average
PM of large numbers of $N$ stars, it is possible to reach the required
accuracies (since random errors scale as $N^{-0.5}$).

The highest-quality optical/near-IR PM measurements are obtained from
space, and specifically with the Hubble Space Telescope (HST). HST's
location above the earth atmosphere yields several advantages,
including high spatial resolution, low sky background, and exquisite
telescope stability. Moreover, there is a large data archive of
observations spanning 24 years (or 12 years when restricted to the
highest quality cameras ACS and WFC3), thus making possible
comparisons over large time baselines. While the field of view of HST
is small compared to what is available from the ground, HST's ability
to observe very faint sources yields more PM measurements per fixed
area. The number of measurable stars in an HST field can range from $N
\approx 10^{1-2}$ for a sparse halo pointing (Deason \etal 2012; Sohn
\etal 2014) to $N \approx 10^{5-6}$ for the center of a globular
cluster (e.g., Anderson \& van der Marel 2010).

Due to HST's unparalleled astrometric capabilities, it can achieve the
PM accuracy needed for LG science {\it for individual stars} ($50 \>
\mu$as/yr corresponds to a motion of $\sim 0.01$ HST CCD pixels over
10 years). So it is possible not only to measure very accurate average
PMs, but also to study internal PM kinematics of stellar systems. This
includes studies of the velocity dispersion of globular clusters (not
discussed in this review, but see, e.g.: McNamara \etal 2003, 2012;
McLaughlin \etal 2006; van der Marel \& Anderson 2010; Bellini \etal
2013), the rotation of galaxies (van der Marel \& Kallivayalil 2014),
or different kinematical populations in tidal streams (Sohn \etal
2014). These possibilities have led several groups over the past
decade to focus on the use of HST for LG PM studies. Specifically,
many of the results discussed in the following sections were obtained
in the context of our HSTPROMO (Hubble Space Telescope Proper Motion)
collaboration, a set of several dozen HST investigations with both
observational and theoretical components (see the review in van der
Marel 2014, or the HSTPROMO web
page\footnote{http://www.stsci.edu/$\sim$marel/hstpromo.html}).

Observationally determined PMs can be of two kinds: relative PMs or
absolute PMs. Relative PMs are easier to obtain, and suffice e.g. if
the goal is to measure the {\it internal} kinematics of a stellar
system. But for LG science it is generally necessary to obtain
absolute PMs, by measuring the motion of foreground stars with respect
to distant (stationary) background objects. Most early studies (e.g.,
Piatek \etal 2002; Kallivayalil \etal 2006a) focused on fields
centered on a single background quasar (see, e.g., Fig.~1). A quasar
has the advantage of being a point source, so that its position can be
measured with the same techniques as the foreground stars (Anderson \&
King 2000). However, this limits the applicability to the small
fraction of fields that contain a known quasar. It also limits the
accuracy per field to the accuracy with which a single source can be
centroided.


\begin{figure}[t]
\null\vskip0.15truecm
\leftline{%
\epsfysize=0.47\hsize
\epsfbox{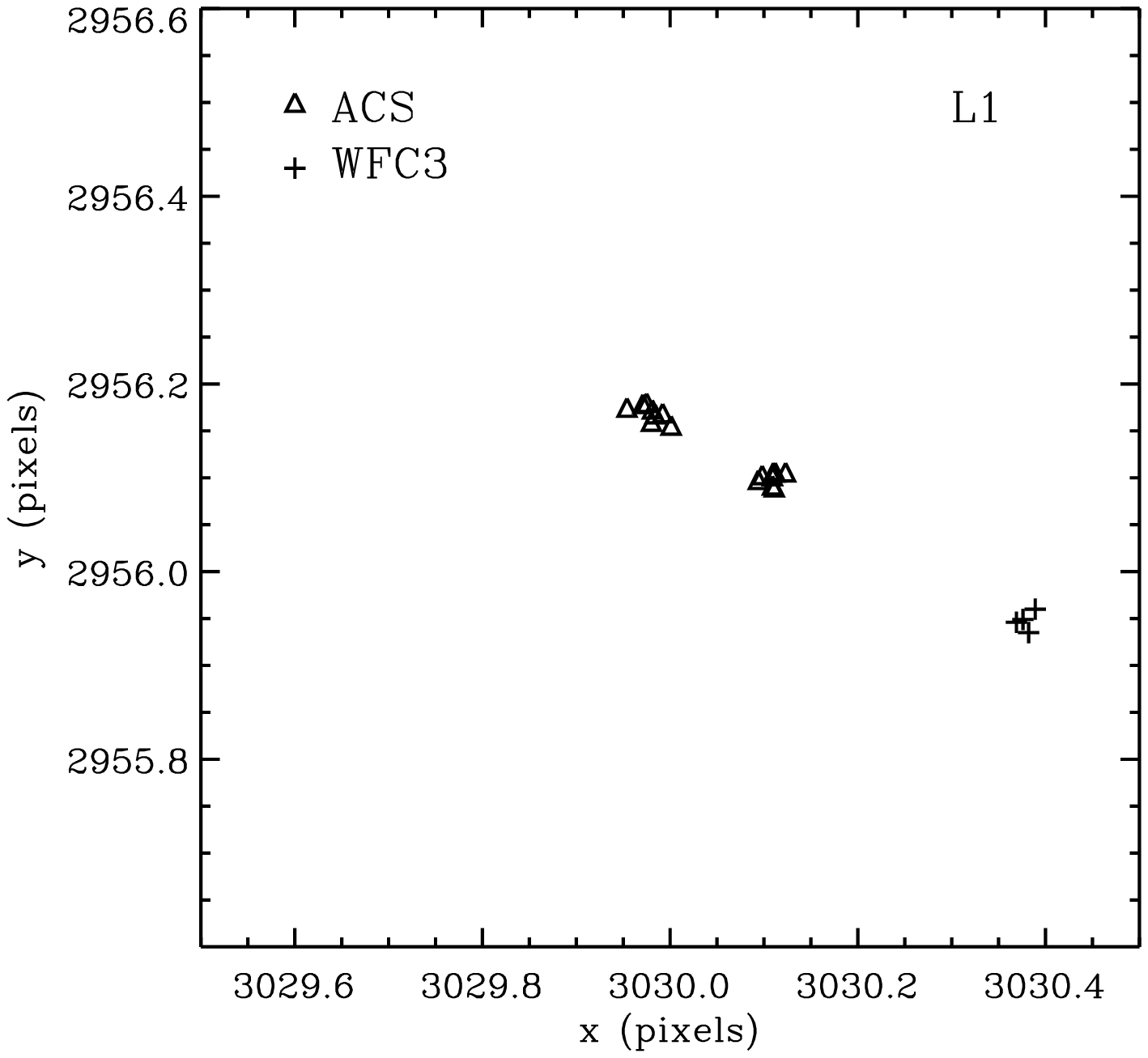}\hfill
\epsfysize=0.47\hsize
\epsfbox{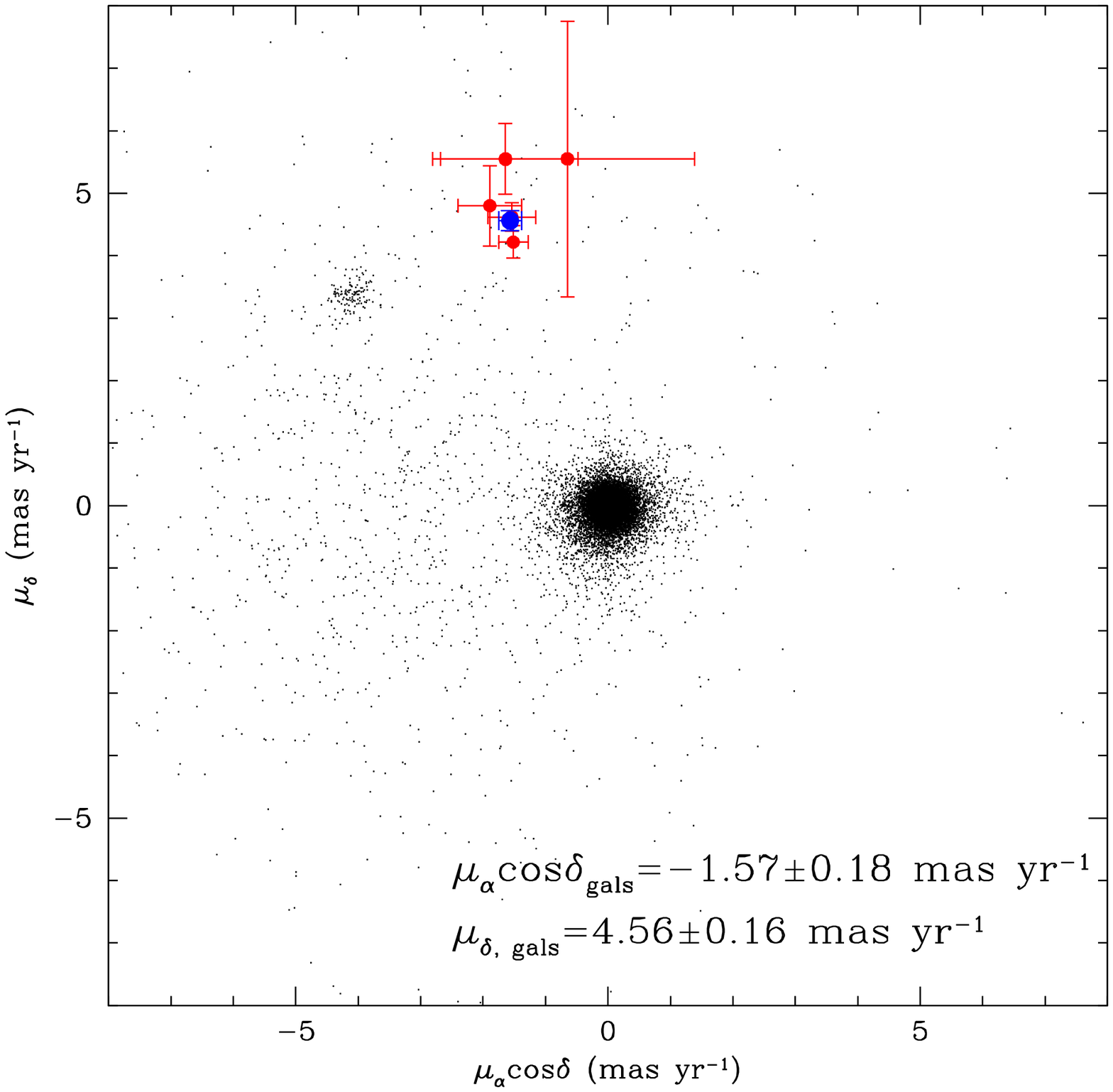}}
\vspace{-0.00truecm} {\small \noindent {\bf Figure 1 (left)} HST
  measurements of the position of a background quasar relative to the
  stars in an LMC field (from Kallivayalil \etal 2013).  The box is
  the size of a single CCD pixel. Measurements are shown with
  different instruments, at three epochs spread over a 7
  year period. The average absolute 
  PM of the LMC stars can be accurately measured from
  such data.\\ \\{\bf Figure 2 (right)} PMs of all compact
  sources in an HST pointing centered on the bulge globular cluster
  M70 (from Massari \etal 2013), measured relative to the cluster
  PM. Point sources (black) are stars in either M70 (middle clump),
  the Sagittarius dSph galaxy in the background (top left clump), or
  the MW disk and bulge (extended distribution left of center). Red
  points with errorbars, and their weighted average in blue, are
  distant background galaxies which define the absolute PM
  zeropoint. Such data allow the measurement of both absolute bulk and 
  relative internal motions of globular clusters and dwarf galaxies.}\\
\end{figure}


It is possible to achieve higher absolute PM accuracies by using
compact background galaxies as stationary reference sources (see,
e.g., Fig.~2). While their individual positional accuracy is somewhat
poorer that for a point source, their much larger number provides an
overwhelming $\sqrt{N}$ advantage in averaging (a typical 1 hour HST
pointing will contain of order $10^2$ background galaxies sufficiently
bright and compact for astrometry). Also, background galaxies sample
the whole detector, which reduces systematic errors upon
averaging. Measuring positions of background galaxies is more
complicated than for point sources, but sophisticated techniques for
this now do exist (e.g., Mahmud \& Anderson 2008; Sohn \etal
2012,2013,2014).

Systematic errors are often the limiting factor in PM studies. Thus it
is necessary to perform careful calibrations of PSF shapes, geometric
distortions, charge-transfer efficiency, color effects, time
variability in calibrations, etc. (e.g., Anderson \& King 2000; Sohn
\etal 2012). When this is done, the systematics can be controlled to
levels below the typical random errors. This has been confirmed for
several studies by the good agreements between HST PM measurements and
independent estimates from combinations of theoretical models and LOS
velocity measurements (e.g., van der Marel \etal 2012a; van der Marel
\& Kallivayalil 2014; Sohn \etal 2014).

The results from GAIA, a full-sky astrometric satellite launched in
2013, will further revolutionize our understanding of PM
dynamics. However, the best results will likely be limited to the MW
and its satellites. This is because GAIA is not optimized for faint
targets or crowded fields. For example, the predicted end-of-mission
accuracy at magnitude $V \sim 20$ is $\Delta$ PM $\sim 430 \>
\mu$as/yr (this includes the contributions from scattered light
discovered post-launch). By comparison, HST already reaches $\Delta$
PM $\sim 100 \> \mu$as/yr at $V \sim 25$ (e.g., Sohn \etal
2014). Therefore, HST will continue to be a unique and complementary
resource in the GAIA era. Looking further into the 2020s, the EUCLID
and WFIRST-AFTA space missions will enable HST-quality PM measurements
over much wider fields of view, thus opening the door for even more
ambitious scientific investigations of the LG.

\section{Milky Way Populations}

\noindent Studies of the dynamics of the MW disk through PM
measurements of stars in the solar neighborhood date back more than a
century. The Hipparcos mission has produced the most important catalog
in this field (e.g., Dehnen \& Binney 1998), and the GAIA mission will
further improve our knowledge. I will not review the local MW disk
dynamics here, but will focus instead on more distant MW components
and tracers.

\smallskip 
\noindent {\bf Globular Clusters\quad} The absolute PMs of many
globular clusters have been measured from ground-based observations in
a series of papers by Casetti-Dinescu \etal (e.g., 2007, 2010,
2013). A catalog with the PMs of $\sim 60$ clusters is available
online\footnote{http://www.astro.yale.edu/dana/gc.html}. The PMs have
been used, e.g., to assess the three-dimensional orbits of clusters,
the equilibrium dynamics of the cluster population as a whole, the
dynamical differences between disk and halo clusters, the possible
external origin of some clusters, or potential associations of
specific clusters with tidal streams. However, the available
ground-based PM uncertainties are such that for $D \gtrsim 10 \kpc$,
the corresponding velocity uncertainties are generally too large to
strongly constrain these questions. More accurate HST PM measurements
of globular clusters are now starting to become available as well
(e.g., Massari \etal 2013; Sohn \etal 2014; see Fig.~2). These have
the potential to probe the cluster dynamics to much larger distances.

\smallskip 
\noindent {\bf Galactic Bulge\quad} Moving on to individual stars
outside the solar neighborhood, several groups have studied the
dynamics of the MW Bulge. Early ground-based PM measurements
(Spaenhauer \etal 1992) found the velocity dispersions to be
near-isotropic.  Subsequent HST studies have revealed more complex
kinematical signatures, such as rotation, and radial (depth)
variations within the bulge (Kuijken \& Rich 2002; Clarkson \etal
2008; Soto \etal 2014). These have been used to improve our
understanding of the bulge's structure and its connection to the
central bar (Rattenbury \etal 2007; Poleski \etal 2013).

\smallskip 
\noindent {\bf Metal-Poor Halo\quad} The PMs of distant stars in the
metal-poor halo can also be measured with HST.  Deason \etal (2013)
showed that it is possible to uniquely identify main-sequence halo
stars in random, multiply-imaged HST fields, by combining PM and
color-magnitude diagram (CMD) information. They measured the PMs of 13
stars at $24 \pm 6 \kpc$ toward the Andromeda-Triangulum halo
overdensity, which imply a halo velocity ellipsoid that is roughly
isotropic. This is more tangential than the velocity ellipsoid of halo
stars near the Sun, which is measured from ground-based PMs to be
highly radially anisotropic (e.g., Bond \etal 2010). These results are
consistent with the hypothesis that the Andromeda-Triangulum
overdensity may be a shell in the halo, possibly resulting from an
ancient accretion event. Work is ongoing to use data for many more
fields in the HST Archive to measure PMs for $\sim 1000$ distant stars
throughout the halo (Sohn et al., in prep.). This will help constrain
halo formation mechanisms (through the implied velocity dispersion
anisotropy of halo stars, and the amount of substructure in velocity
space), and it will improve our knowledge of the radial MW mass
profile by resolving the mass vs.~velocity dispersion anisotropy
degeneracy.

\smallskip 
\noindent {\bf Hypervelocity Stars\quad} Dynamical interactions near
the Milky Way's central supermassive black hole can eject stars at
high velocities. Such stars are observed as hypervelocity stars in the
MW halo (Brown \etal 2005). Given their distances, only HST has so far
been able to measure their PMs. An accurate PM can confirm the origin
near the Galactic Center (Brown \etal 2010), while the implied
deviation from purely radial motion can in principle constrain the
shape of the MW's dark halo (Gnedin \etal 2005).

\smallskip 
\noindent {\bf Tidal Streams\quad} The MW halo contains many stellar
streams, composed of the tidally stripped material from globular
clusters or dwarf satellites. These streams hold great promise to
constrain the shape and density profile of the MW's dark matter halo,
especially when PMs are available in addition to sky positions,
distances, and LOS velocities. Koposov \etal (2010) used ground-based
USNO-B astrometry to study the PMs of the GD1 Stream. However, since
this stream is at $D \approx 10 \kpc$, it provides only limited
information on the MW's dark matter halo. The Sagittarius Stream,
composed of material stripped from the Sgr dSph, is a more useful
tracer in this regard, since it reaches distances up to $\sim 100
\kpc$ (Belokurov \etal 2014). For the Sgr dSph itself, both
ground-based (Dinescu \etal 2005) and HST-based PMs (Pryor \etal 2010;
Massari \etal 2013; Sohn \etal 2014; see Fig.~2) are available. For
the Sgr Stream, Carlin \etal (2012) and Koposov \etal (2013) obtained
ground-based PMs along the trailing arm, while Sohn \etal (2014)
obtained higher accuracy HST PMs along both arms. The latter study was
even able to kinematically separate trailing-arm from leading-arm
stars {\it within the same field}, based on their PMs. The average
stream PMs are close to the predictions of the best-fitting Stream
model previously constructed by Law \& Majewski (2010). In this model,
the MW dark halo is near-oblate and oriented orthogonal to the MW
disk. Further measurements and models will likely be required to
either validate or refute this model (see also, Deg \& Widrow 2013;
Ibata \etal 2013; Vera-Ciro \& Helmi 2013). The PMs of other streams,
including the Orphan stream, are also being studied with HST (van der
Marel et al., in prep.).
   
\section{Dwarf Satellite Galaxies}

\noindent The PMs of almost all of the MWs classical dSph satellites
have been measured using HST (Piatek \etal 2003, 2005, 2006, 2007;
Lepine \etal 2011; Pryor \etal 2014), and these are generally the most
accurate measurements available. Knowledge of the PMs has been used to
determine the dSph orbits. These can be used to investigate the origin
of these galaxies, the cause of structural of star formation history
features, and/or their connection with other satellites. For example,
there has long been evidence that some MW satellites lie in a plane
(e.g., Lynden-Bell \& Lynden-Bell 1995). By measuring PMs it is
possible to verify that the orbital angular momentum vectors are
roughly perpendicular to this plane (e.g., Pawlowski \& Kroupa 2013),
as expected for a long-lived configuration. Also, the satellite orbits
can be modeled as an ensemble under the assumption of hydrostatic
equilibrium, to constrain the MW mass profile (e.g., Watkins \etal
2010).

Sohn \etal (2013) measured the absolute PM of Leo~I. This is one of
the most unusual MW satellites, because of its large Galactocentric
distance ($r=261 \pm 13 \kpc$) and the fact that it is rapidly
receding away from the MW ($v_{\rm rad} = 168 \pm 3 \kms$). The
measured PM implies a significant tangential velocity $v_{\rm tan} =
101 \pm 34 \kms$, indicating that it is not on a radial orbit due to a
slingshot interaction with another satellite. Orbit integrations
indicate instead that it likely entered the MW virial radius for the
first time only $2.3 \pm 0.2 \Gyr$ ago, and had its first pericenter
passage $1.0 \pm 0.1 \Gyr$ ago at $91 \pm 36 \kpc$. Comparison of the
properties of Leo~I to the properties of subhalos in cosmological
simulations yields important new constraints on the mass of the MW
(Boylan-Kolchin \etal 2013). Assuming that Leo~I is the least-bound
classical MW satellite, the implied MW virial mass is $M_{\rm vir}
\approx (1.6 \pm 0.3) \times 10^{12} \Msun$; for lower MW masses it
would be vanishingly unlikely to find a subhalo as distant and rapidly
moving as Leo~I. HST studies of the PMs of other distant dwarf
galaxies are currently ongoing, including both distant MW satellites
such as Leo T (Do et al., in prep.), as well as distant LG satellites
such Cetus, Tucana, Leo A, and the Sgr dwarf Irregular galaxy (Sohn et
al., in prep.).

Future studies, either with HST or future facilities, hold the promise
to determine also the {\it internal} kinematics of dwarf galaxies.
Such measurements have the potential to break the mass vs.~velocity
dispersion anisotropy degeneracy. This can help constrain the nature
of dark matter, by determining whether dark halo mass profiles have
central cores or cusps (Strigari \etal 2007).

\section{The Magellanic Clouds}

\noindent One of the first crude measurements of the PMs of the
Magellanic Clouds was obtained by the Hipparcos satellite (Kroupa \&
Bastian 1997). With HST it has been possible to significantly refine
the accuracy of these measurements. Kallivayalil \etal (2006a,b)
presented PMs based on two epochs of HST data with a 2-year time
baseline. These measurements were recently refined by Kallivayalil
\etal (2013) using a third-epoch of HST data that extends the time
baseline to 7 years (see Fig.~1). The key finding from this work has
been that the Magellanic Clouds move faster about the Milky Way than
was previously believed based on models of the Magellanic Stream.  In
these models, the Clouds were assumed to have been long-term MW
satellites moving on near-circular orbits in a logarithmic potential
(e.g., Gardiner \& Noguchi 1996). By contrast, orbit calculations
based on the HST PMs in a cosmologically motivated MW halo potential
imply either a much longer-period more-elliptical orbit, or that the
Clouds are on their first infall into the MW (Besla \etal 2007). Given
the individual LMC and SMC PM measurements, the first-infall models
are the only ones in which the LMC and SMC form a bound pair
(Kallivayalil \etal 2013), consistent with their structures and star
formation histories.  The plausibility of a first-infall scenario is
further supported by both cosmological simulations (Boylan-Kolchin \etal
2011) and the gas-rich nature of the Clouds (van den Bergh 2006). In a
first-infall scenario, the Magellanic Stream may have formed purely
from interaction between the Clouds, without assistance from MW tidal
or ram pressure forces (Besla \etal 2010; Peebles \& Tully
2013). These results have not just refined our understanding of the
Magellanic Clouds, but also of the formation of Magellanic Irregular
galaxies in general (Besla \etal 2012).

Almost all of our knowledge of the rotation of galaxies is based on
LOS velocity measurements. However, PM measurements can also probe
this rotation. By mapping variations in the HST PM measurements across
the face of the LMC (Kallivayalil \etal 2006a; Piatek \etal 2008), it
has been possible to measure PM rotation for the first time for any
galaxy. van der Marel \& Kallivayalil (2014; see Fig.~3) measured both
the PM rotation field and rotation curve of the LMC in detail, and
demonstrated that the results are consistent with knowledge from LOS
velocity measurements for individual stars. They used the
three-dimensional dynamical information obtained from combination of
the different techniques to obtain new insights into the geometry,
structure, and distance of the LMC. HST studies of the internal PM
dynamics of the SMC (Kallivayalil et al., in prep.) and the Magellanic
Bridge (van der Marel et al., in prep.) are currently ongoing.


\begin{figure}[t]
\null\vskip0.15truecm
\leftline{%
\epsfysize=0.42\hsize
\epsfbox{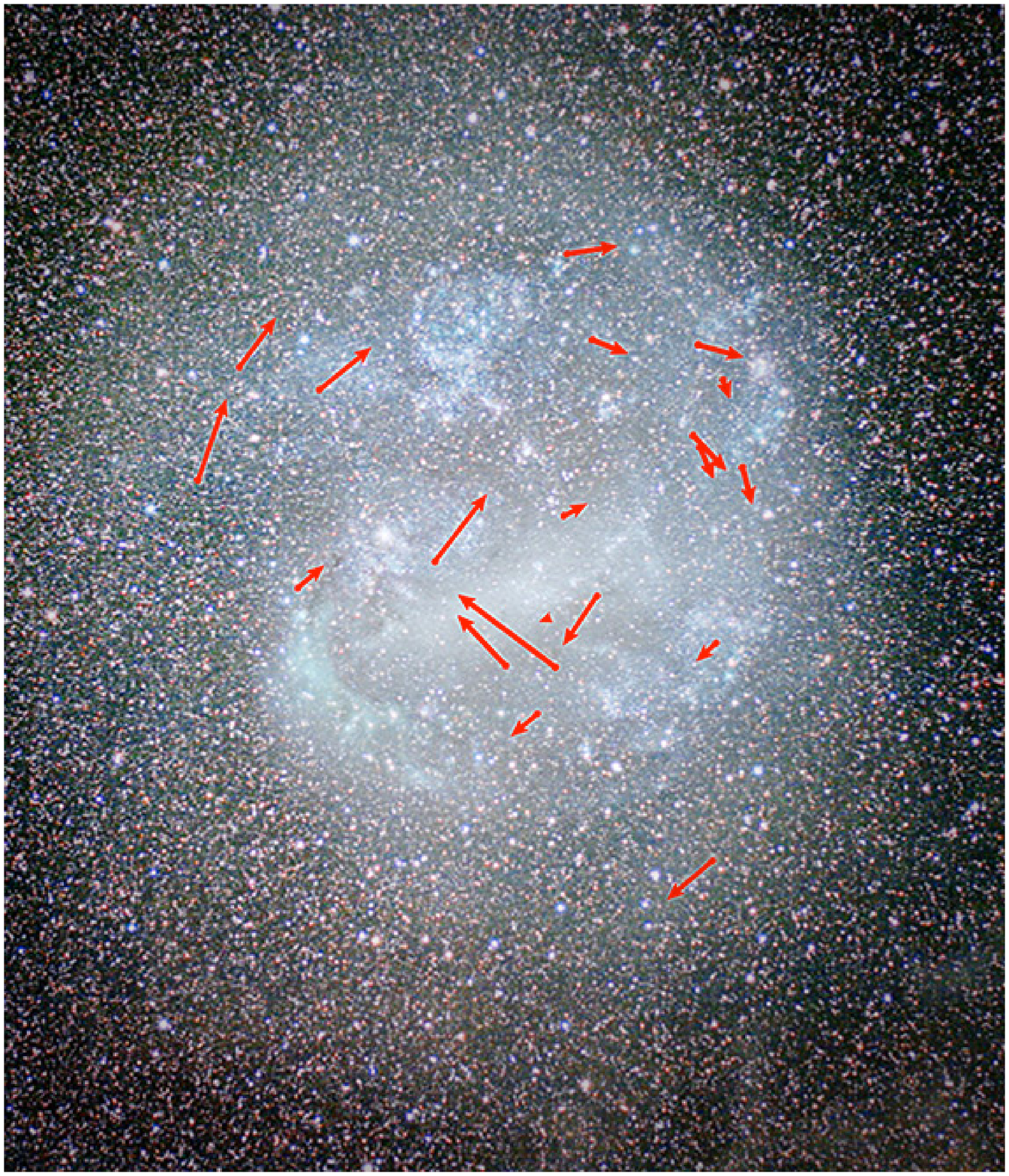}\hfill
\epsfysize=0.42\hsize
\epsfbox{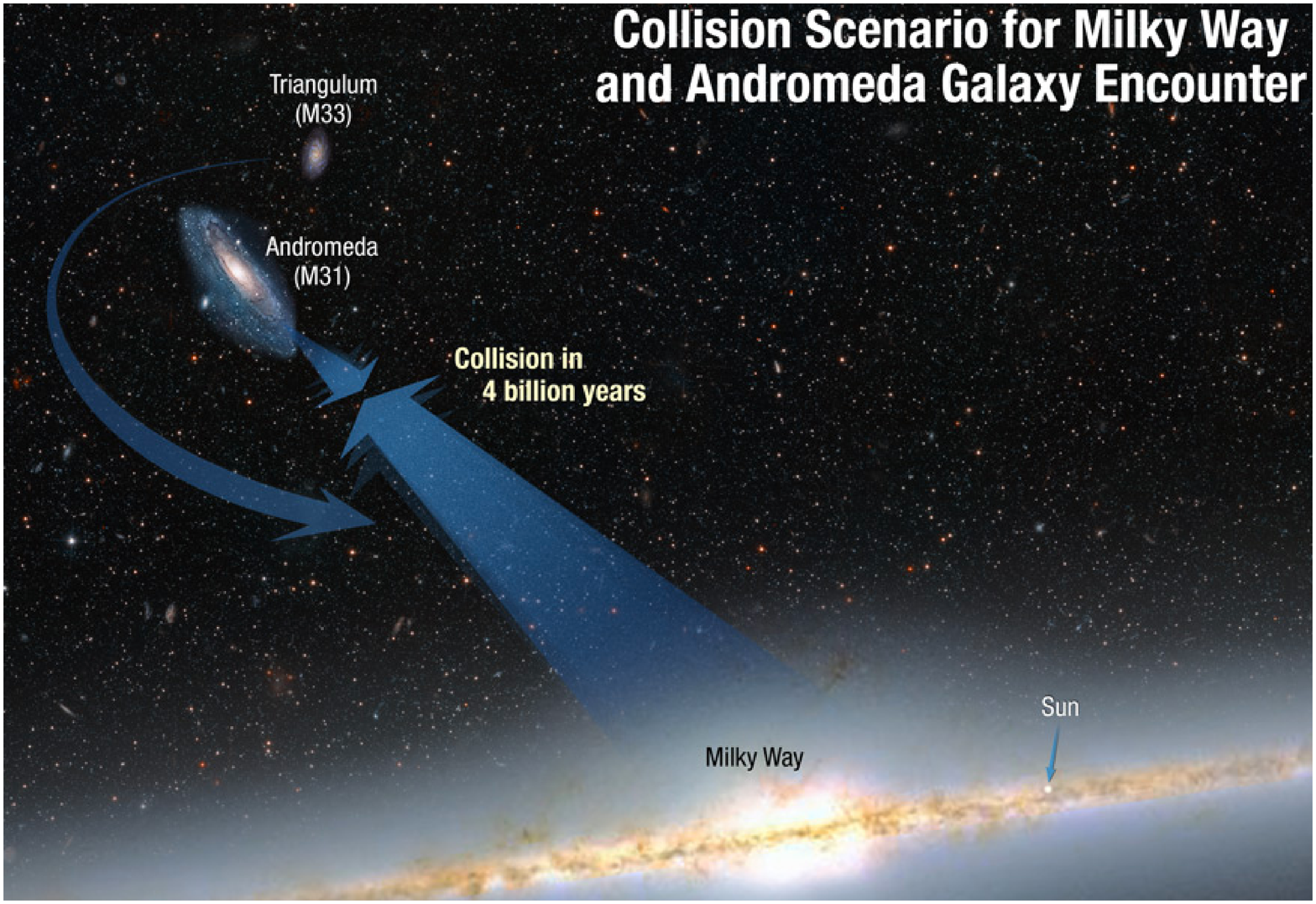}}
\vspace{0.20truecm} {\small \noindent {\bf Figure 3 (left)} Deep
  ground-based LMC image (courtesy Y.~Beletsky, vertical scale $\sim
  14^{\circ}$), with HST measurements of the LMC PM rotation overlaid
  (red arrows; from van der Marel \& Kallivayalil 2014). The LMC makes
  one revolution every few-hundred million years. The PM data can
  measure this rotation accurately, which (when combined with LOS
  data) provides a fully three-dimensional probe of the internal disk
  kinematics.\\ \\{\bf Figure 4 (right)} Illustration of the predicted
  future evolution of the LG, based on the observed M31 PM from HST
  (Sohn \etal 2012) and the observed M33 PM from water maser
  observations (Brunthaler \etal 2005). The MW and M31 will collide in
  $\sim 4 \Gyr$, and they will fully merge in $\sim 6\Gyr$. M33 will
  swing around M31 and take part in the collision as well (from van
  der Marel \etal 2012b).}\\
\end{figure}


\section{The Andromeda System}

\noindent M31 was the first galaxy for which a LOS velocity was
measured, and it has been known for about a century that it is moving
towards the MW (Slipher 1913). However, to understand the future orbit
and evolution of the M31-MW system, and hence the LG, it is necessary
to also know the PM of M31. Sohn \etal (2012) used HST to obtain the
first-ever PM measurement for M31, using data with 5--7 year time
baselines for three different fields studied with multiple HST
cameras. The result agrees with indirect estimates based on the LOS
kinematics of M31 and LG satellites (van der Marel \& Guhathakurta
2008). The combined results imply that the M31 velocity vector is
statistically consistent with a radial (head-on collision) orbit
towards the Milky Way ($V_{\rm tan,M31} \leq 34.3 \kms$ at 1$\sigma$
confidence; van der Marel \etal 2012a).

The M31 and MW protogalaxies started moving away from each other after
the Big-Bang, but their mutual gravity caused the turn-around that led
to their current approach. Their combined mass implied by this ``LG
timing argument'', including corrections for cosmic bias and scatter,
is $M_{\rm LG} = M_{\rm MW, vir} + M_{\rm M31, vir} = (4.93 \pm 1.63)
\times 10^{12} \Msun$. Bayesian combination with other mass estimates
for the individual MW and M31 galaxies yields the more accurate
estimate $M_{\rm LG} = (3.17 \pm 0.57) \times 10^{12} \Msun$ (van
der Marel \etal 2012a).

$N$-body and Monte-Carlo simulations can be used to predict the future
evolution of the MW-M31 system (van der Marel \etal 2012b; see
Fig.~4). The galaxies will merge $t = 5.86^{+1.61}_{-0.72} \Gyr$ from
now. The first pericenter occurs at $t = 3.87^{+0.42}_{-0.32} \Gyr$,
at a pericenter distance $r = 31.0^{+38.0}_{-19.8} \kpc$. In 41\% of
Monte-Carlo orbits M31 makes a direct hit with the MW, defined here as
a first-pericenter distance less than 25 kpc. The PM of the Triangulum
galaxy, M33, is known from water maser observations (Brunthaler \etal
2005), so its orbit can be calculated as well. The most likely outcome
is for the MW and M31 to merge first, with M33 settling onto an orbit
around them that may decay towards a merger later. However, there is a
9\% probability that M33 makes a direct hit with the MW at its first
pericenter, {\it before} M31 gets to or collides with the MW. Also,
there is a 7\% probability that M33 gets ejected from the Local Group,
temporarily or permanently. The radial mass profile of the MW-M31
merger remnant is significantly more extended than the original
profiles of either the MW or M31, and suggests that the merger remnant
will resemble an elliptical galaxy. The Sun will most likely ($\sim
85$\% probability) end up at larger radius from the center of the
MW-M31 merger remnant than its current distance from the MW center,
possibly further than $50 \kpc$ ($\sim 10$\% probability). There is a
$\sim 20$\% probability that the Sun will at some time in the next $10
\Gyr$ find itself moving through M33 (within $10 \kpc$), but while
dynamically still bound to the MW-M31 merger remnant.

\section{Concluding Remarks}

\noindent The ability to measure the PMs of stars and galaxies in the
nearby Universe with HST and other facilities has led to a revolution
in our understanding of the dynamics of the LG and its galaxies. The
LG continues to evolve, with Leo I, and probably also the Magellanic
Clouds, just now falling into the MW for the first time. In a few
billion years, M31 will arrive as well, which will cause the Local
Group to be drastically reshaped. With the continuing improvements in
observational facilities, and the prospect of GAIA and other future
space telescopes, it is likely that PM measurements will continue to
improve our understanding of our dynamic LG environment. In turn, this
will be a key ingredient for understanding the formation and evolution
of galaxies and galaxy groups in general.

\section*{Acknowledgements}

\noindent Support for this work and for HSTPROMO projects was provided
by NASA through grants from STScI, which is operated by AURA, Inc.,
under NASA contract NAS 5-26555.

\end{document}